 \documentclass[12pt]{article}
 \usepackage{graphicx}
 \usepackage{amssymb}

\usepackage{amssymb}

\def\'#1{{\accent19\ifx #1i \i\else #1\fi}}

\def\be{\begin{equation}}
\def\ee{\end{equation}}
\def\bea{\begin{eqnarray}}
\def\eea{\end{eqnarray}}

\newcommand{\boldmathnabla}  {\mbox{$\nabla$}}

\newbox\Ancha
\catcode`@=11
\newdimen\ex@
\title{Exact solution for charged-particle propagation
      during a first-order electroweak phase transition with 
       hypermagnetic fields}
 \author{Jaime Besprosvany$^\dagger$ and Alejandro
 Ayala$^\ddagger$}
\date{  $^\dagger$Instituto de F\'{\i}sica,
         Universidad Nacional Aut\'onoma de M\'exico
         Apartado Postal 20-364,
         M\'exico Distrito Federal 01000, M\'exico\\
         $^\ddagger$Instituto de Ciencias Nucleares, Universidad Nacional
         Aut\'onoma de M\'exico, Apartado Postal 70-543,
         M\'exico Distrito Federal 04510, M\'exico}
          
\begin{document}
          \maketitle
\begin{abstract}

We obtain the exact solution of the Klein-Gordon equation
describing the propagation of a particle in two regions of
different constant magnetic field, separated by an infinite plane
wall. The continuity of the wave function and of its derivative at
the interface is satisfied when including evanescent-wave terms.
We analyze solutions on truncated spaces and compare them with
previously obtained approximate solutions. The findings of this
work have applications in the problem of the propagation of
particles in the presence of a bubble wall in the midsts of an
electroweak phase transition, where the two regions separated by
the wall are influenced by different (hyper)magnetic field
strengths.

\end{abstract}

\centerline{12.15.Ji, 03.65.Pm, 98.80.Cq, 98.62.En}

\maketitle

\section{Introduction}\label{I}

The possible existence of magnetic fields in the early universe
has recently become the subject of intense research due to the
various interesting cosmological implications~\cite{reviews,
Fogli}. Among them is the effect that primordial magnetic fields
could have had on the dynamics of the electroweak phase transition
(EWPT) at temperatures of order $T\sim 100$ GeV~\cite{asps}. It
has been recently pointed out that, provided enough {\it CP}
violation exists, large scale primordial magnetic fields can be
responsible for a stronger first-order
EWPT~\cite{{Giovannini},{Elmfors}} in analogy with a type I
superconductor where the presence of an external magnetic field
modifies the nature of the superconducting phase transition due to
the Meissner effect. Primordial magnetic fields, extending over
horizon distances, can be generated for example in certain types
of inflationary models through the breaking of conformal
invariance~\cite{magneticinfla}.

Recall that for temperatures above the EWPT, the
SU(2)$_L\times$U(1)$_Y$ symmetry is restored and the propagating,
non-screened vector modes that represent a magnetic field
correspond to the U(1)$_Y$ group, instead of  the U(1)$_{em}$
group, and are therefore properly called {\it hypermagnetic}
fields.

Another interesting implication of the presence of these fields
during a first-order EWPT concerns the scattering of fermions off
the true vacuum bubbles nucleated during the phase transition. In
fact, using a simplified picture of a first-order EWPT, it has
been shown~\cite{Ayala2} that the presence of such fields also
provides a mechanism, working in the same manner as the existence
of additional {\it CP} violation within the SM, to produce an
axial charge segregation. The asymmetry in the scattering of
fermion axial modes is a consequence of the chiral nature of the
fermion coupling to hypermagnetic fields in the symmetric phase.

The simplified treatment of the scattering process allows to
formulate the problem in terms of solving the Dirac equation with
a position-dependent fermion mass~\cite{Ayala}. This dependence is
due to the fact that fermion masses are proportional to the vacuum
expectation value of the Higgs field, this last being zero in the
false phase and non-vanishing in the broken symmetry phase.
Transmission and reflection coefficients can be computed
approximately by looking into the most likely transitions from a
given incident flux of particles.

Nevertheless, in spite of the success of such a scheme, it is
clear that a more accurate account of such processes is needed in
order to set better limits that constrain the feasibility of the
mechanism for axial charge segregation~\cite{AyaBespro}.

In this work, we present an exact solution to the problem of
describing the propagation of a particle in two regions of
different constant magnetic field, separated by an infinite plane
wall, corresponding to the wall of a bubble nucleated during a
first-order EWPT. The particle is massless outside the bubble, in
the symmetric phase, and massive inside the bubble, in the
broken-symmetry phase. The analysis is made in the infinitely
thin-wall approximation to avoid  unnecessary numerical
complications. In physical terms, this means that the spatial
region over which particle masses change is small compared to
other relevant length scales such as the particle mean free path.
The choice of external field captures such information, namely,
changes in the magnetic field occur over large spatial regions,
and thus can be considered homogeneous near the interphase.

Finite temperature effects are implicitly included in the
treatment of the problem. One could model the change of the
particle's mass with position using the so-called {\it kink}
solution for the spatial profile of the Higgs field~\cite{Ayala}.
This profile is obtained from the classical solution to the
equations of motion with the finite-temperature effective
potential. The wall profile considered in this work can be thought
of as obtained from the kink solution in the limit where the wall
thickness goes to zero.

In order to provide a working solution, we perform a numerical
analysis by truncating the solution space. To present the method,
we choose to work within the framework of the Klein-Gordon
equation although applications to the Dirac equation are
straightforward.

The outline of this work is as follows: In Sec.~\ref{II}, we
briefly review the well-known solution to the problem of a
spinless, charged particle moving in a constant magnetic field. In
Sec.~\ref{III}, we solve the problem that describes the motion of
this particle propagating through a planar interface dividing two
regions with different magnetic field strengths, setting the
system of equations that allows to find its reflection and
transmission amplitudes. In Sec.~\ref{IV} we find a numerical
solution by truncating the infinite tower of coupled equations
that result from the exact treatment. In Sec.~\ref{V}, we use the
above solutions to compute reflection and transmission
coefficients for modes moving from the symmetric phase toward the
broken-symmetry phase and explicitly check that these satisfy flux
conservation. In Sec.~\ref{VI} we build a one-dimensional model to
understand such flux conservation in spite of the relativistic
context of the problem. In Sec.~\ref{VII} we summarize this work
and give an outlook about its implications. Finally, in the
appendix we   explain how the proposed method is applied for
arbitrary constant magnetic fields.

\section{Klein-Gordon equation and solutions for constant magnetic
         field}\label{II}

We start by describing the motion of a particle with charge $e_0$
and mass $\mu$, under a  constant magnetic field of magnitude $B$.
Working in a gauge where the time component of the corresponding
vector potential is zero, we can choose
\begin{eqnarray}
   \label{3dmag}
   {\bf A} =-\frac{1}{2} {\bf r} \times {\bf  B},
\end{eqnarray}
with ${\bf B}=(0,0,B)$. By minimal substitution ${\bf
p}\rightarrow {\bf p}+e_0{\bf A}$ in the Klein-Gordon equation we
have
\begin{eqnarray}
   \label{equMAg3mod}
   [-{\boldmathnabla}^2
   + \frac{1}{4}e_0 ^2 B^2 r^2+e_0 {B} L_z + \mu^2]
   \Psi=E^2 \Psi,
\end{eqnarray}
where we use $\hbar=c=1$ and, working in cylindrical coordinates
$(r,\phi,z)$, $r=\sqrt{x^2+y^2}$ is the radial coordinate. Also,
$L_z=-i(\partial_y x-
\partial_x y)$ is the  $z$-component of the orbital angular
momentum.

The solutions are given in various representations  in
Ref.~\cite{Sokolov},\cite{JaiMatias}. In one  choice, a complete
set of eigenfunctions is characterized by an angular momentum
quantum number $l\in {\mathbb Z}$, and we assume $l\geq 0$; a
radial transverse $xy$-plane excitation quantum number
$s=0,1,2...$; and  free-particle propagation along the
magnetic-field direction, with wave number $k_n$, which with our
choice of magnetic field direction as in Eq.~(\ref{3dmag}), points
along the $\hat {\bf z}$ axis. The energy is given by
\begin{eqnarray}
   \label{massdecoupledeqs}
   E=\sqrt{k_n^2+2e_0B( n+1/2)+\mu^2},
\end{eqnarray}
and the eigenfunctions by
\begin{eqnarray}
   \label{eigenfunctionsCompl}f_B^{ns}(r,\phi,z)= \frac{1} {L^{1/2}(2
   \pi)^{1/2}}N_{ns} I_{ns}(\rho)e^{ik_nz}e^{il\phi},
\end{eqnarray}
with
\begin{eqnarray}
   \label{eigenfunctionsIns}
   I_{ns}(\rho)=\frac{1}{(n!s!)^{1/2}}\rho^{l/2}e^{-\rho/2}
   Q^l_s(\rho),
\end{eqnarray}
where $N_{ns}=(e_0B/E)^{1/2}$, $\rho=(e_0B/2)r^2$,  $n=s+l$,  $L$
is the distance along $z$ and it eventually cancels  out in the
calculation of reflection and transmission coefficients. The
definition of $N_{ns}$ differs from that of Ref.~\cite{Sokolov} by
a $\mu^{1/2}$ factor. For $k_n$ real, the sign of the exponent in
$e^{ik_nz}$ represents the direction of motion of the wave. We
note that the set of quantum numbers $(n,s,k_n)$, $(n,l,k_n)$,
$(s,l,k_n)$ can be used interchangeably.

The normalization coefficients are such that the set of
eigenfunctions satisfy the normalization condition
\begin{eqnarray}
   \label{normTot}
    \int  d^3x \rho_0 =  E\int d^3x |f_B^{ns}|^2=1,
\end{eqnarray}
where $\rho_0$ is the charge density. $Q^l_s(\rho)$ is a Laguerre
polynomial defined by
\begin{eqnarray}
   \label{LaguerreSok}
 Q^l_s(\rho)=s!  L^l_s(\rho),
\end{eqnarray}
where we use the convention~\cite{Abramowitz}
\begin{eqnarray}
   \label{LaguerreConv}
   L^l_s(\rho)= \sum_{j=0}^s
   \frac{(s+l)!\rho^{s-j}}{j!(s-j)!(s+l-j)!}.
\end{eqnarray}
For negative $l$, the solutions are
$Q^l_s(\rho)=(-1)^l\rho^{-l}Q^{-l}_{s+l}(\rho),$ and thus the
energy eigenvalue is $E=\sqrt{k_n^2+2e_0B( s+1/2)+\mu^2}$.
Correspondingly, a new complete basis set over $s$ on the $xy$
plane can be defined.

A general solution $\Psi_{El}$ can be constructed from a linear
superposition of the basis of solutions in
Eq.~(\ref{eigenfunctionsCompl}) and is given by
\begin{eqnarray}
   \label{GenSol}
   \Psi_{El}(r,\phi,z) = \sum_{n} C_n
   f_B^{ns}(r,\phi,z),
\end{eqnarray}
where $C_n$ are arbitrary coefficients that weigh the
corresponding Landau-level contributions.

All such modes are in principle necessary in the context of
scattering, where an incoming wave is either reflected or
transmitted conserving energy, but  the magnitude of its momentum
eigenvalue along the $\hat{z}$ axis, $k_n$, may change.  Indeed,
the  energy gained or lost along the $\hat{z}$ axis is
redistributed in the form of radial excitations associated with
the quantum number $s$. Thus, changes in $s$ are compensated by
changes in $k_n$  in such a way that the energy is conserved, and
is equal to the one in Eq.~(\ref{massdecoupledeqs}). Notice that,
for large enough $n$ and  fixed energy $E$, $k_n$ will eventually
become imaginary. We refer to this kind of solution as   {\it
evanescent} wave. Physically, this situation mimics the
transmitted wave function found in the presence of a square
barrier with a potential value larger than the energy $E$.

That momentum along the $\hat{z}$ axis is not necessarily
conserved is physically clear from the fact that translational
invariance along the $\hat{z}$ axis in this problem is lost due to
the presence of the wall, as will be exemplified in the
one-dimensional models in  Sec.~\ref{VI}.

\section{Solution for two regions with different magnetic fields
         separated by an infinite plane wall}\label{III}

We proceed to find the solution for the propagation of a particle
in two regions with different constant magnetic fields, and
different mass. We use the solutions found in the previous
section, placing our coordinate system with the $xy$ plane on the
wall separating the two regions. For definiteness, we call region
$I$ the half-space with $z<0$, where the field strength is $B$,
the coupling constant $e_0$, and $\mu=0$, and region $II$ the
half-space with $z>0$, where the field strength is $B^\prime$, the
coupling constant $e_0'$, and $\mu\neq 0$. The magnetic fields
keep perpendicular to the wall.

An incoming flux of charged, spinless and massless particles from
region $I$, propagating along the $z$ axis, scatters off the wall
at $z=0$, and is reflected back into region $I$, and transmitted
into region $II$.

We prepare the incoming flux with   each  particle propagating
with momentum $k_m^i$,   quantum number  $s$ --representing a
given transverse oscillation mode-- and angular momentum $l$.
These values define $E$. The particle's wave function satisfies
the Klein-Gordon equation with magnetic fields $B$ and $B'$,
corresponding to  regions $I$ and $II$, respectively. Notice that
the energy and angular-momentum component along the $z$ direction
are conserved. This means that $E$ and $l$ will be good quantum
numbers on both sides of the wall. Also, we demand that the wave
function and its derivative be continuous at the interface.

We normalize the incoming wave function to unity, as in
Eq.~\ref{normTot}, which from Eq.~(\ref{GenSol}) means that
$C_m^I=1$. Among the scattered waves, one corresponds to a
reflected wave with magnitude of the momentum $k_m$ equal to the
incident one, but with direction of motion reversed
\begin{eqnarray}
   \label{refkm}
   k_m^i=-k_m= \sqrt{E^2-2e_0B( m+1/2)} \, ;
\end{eqnarray}
we associate to it the coefficient $C^{I}_m$, which represents the
reflection amplitude. The total wave function in region $I$ can be
expressed, according to Eq.~(\ref{GenSol}), as
\begin{eqnarray}
   \label{GenSolI}\Psi_{El}^I(r,\phi,z)= f_{B}^{m s}(r,\phi,0)
   e^{ik_{m}^i z}+C^{I}_m f_{B}^{m s}(r,\phi,0) e^{i k _{m}
   z}+\sum_{n\neq m} C_n^I f_B^{ns}(r,\phi,z),
\end{eqnarray}
where in the first term we have used that
\begin{eqnarray}
   \label{expansionWFnonPrime}
   f_{B}^{ns}(r,\phi,z)=f_{B}^{n  s }(r,\phi,0)
   e^{ik_{n} z},
\end{eqnarray}
as implied by Eq.~(\ref{eigenfunctionsCompl}).

Recall that given an energy eigenvalue $E$ and an orbital angular
momentum $l$, to each eigenfunction--labeled by $n$--among the
infinite set, corresponds a value of the momentum $k_n$. Thus,
from Eq.~(\ref{massdecoupledeqs}), the possible values for $k_n$
are given by
\begin{eqnarray}
   \label{refk} k_n = \left \{
   \begin{array}{cc}
      -\sqrt{E^2-2e_0B( n+1/2)},  \ \ \  E^2 \ge 2e_0B( n+1/2)   \\
      -i\sqrt{-E^2+2e_0B( n+1/2)},\ \ \  E^2 < 2e_0B( n+1/2).
   \end{array}        \right .
\end{eqnarray}
Notice that the majority of the $k_n$ are imaginary, for there is
a value $n$ beyond which the square-root argument is negative. The
sign of  the imaginary $k_n$ is chosen so as to have a finite wave
function as $z\rightarrow-\infty$.

Also, a general solution in region $II$ can be written as
\begin{eqnarray}
   \label{SolII}
   \Psi_{El}^{II}(r,\phi,z)= \sum_{n^\prime}
   C_{n^\prime}^{II}
   f_{B^\prime}^{{n^\prime}s^\prime}(r,\phi,z),
\end{eqnarray}
where the solutions have the same $l$ and $E$, and a quantum
number $n'$ given by $n^\prime=s^\prime+l$, with a value of the
longitudinal momentum, chosen  to represent  transmitted  or
tunneled evanescent waves, given by
\begin{eqnarray}
   \label{refkprime}
   k_{n^\prime} = \left \{
   \begin{array}{cc}
      \sqrt{E^2-2e_0'B^\prime( n^\prime+1/2)-\mu^2},
      \ \ \  E^2 \ge 2e_0'B'( n^\prime+1/2)+\mu^2   \\
      i\sqrt{-E^2+2e_0'B^\prime( n^\prime+1/2)+\mu^2},\ \ \  E^2 < 2e_0'B'(
      n^\prime+1/2)+\mu^2,
   \end{array}     \right .
\end{eqnarray}
where the sign of a real $k_{n^\prime}$ represents a wave
travelling in the $+z$ direction, and the sign of an imaginary
$k_{n^\prime}$ is chosen so as to have a finite wave function for
$z\rightarrow \infty$.

The constants $C^{I}_m$, $C_{n}^{I}$, $C_{n^\prime}^{II}$ in
Eqs.~(\ref{GenSolI}) and~(\ref{SolII}) are coefficients to be
determined from the continuity of the solution and its derivative
across the planar interphase.

First, in order to join the solutions from regions $I$ and $II$,
we can re-express the region-$II$ transverse wave functions in
terms of the transverse-component basis of the region-$I$
solutions  at $z=0$:
\begin{eqnarray}
   \label{expansionWFbarrier} f_{B^\prime}^{n^\prime
   s^\prime}(r,\phi,0)=\sum_{s}a_{ s s^\prime}^l f_{B}^{ns}(r,\phi,0),
\end{eqnarray}
where the $z$ dependence factors out as in
Eq.~(\ref{expansionWFnonPrime})
\begin{eqnarray}
   \label{expansionWF}
   f_{B^\prime}^{n^\prime s^\prime}(r,\phi,z)=f_{B^\prime}^{n^\prime
   s^\prime}(r,\phi,0) e^{ik_{n^\prime} z}.
\end{eqnarray}
The amplitude $a_{ s s^\prime}^l$ can be calculated in a closed
form as
\begin{eqnarray}
   \label{amplitudBBp}
   a_{ s s^\prime}^l = \frac{1}{2}N_{ns} N_{n^\prime s^\prime}
   \left(\frac{s! s^\prime!}{n! n^\prime! }\right)^{1/2}(e_0e_0'B
   B^\prime/4)^{l/2}\int_0^\infty dx e^{-bx}x^l L^l_s(\lambda x)
   L^l_{s^\prime}(\eta x),
\end{eqnarray}
where we  have changed    variable with  $x=r^2$, and have carried
out the angular integration; we have also defined
$\lambda=e_0B/2,$ $\eta=e_0'B^\prime/2$,
$b=(e_0B+e_0'B^\prime)/4$. The integral in Eq.~(\ref{amplitudBBp})
is given by~\cite{Gradshtein}
\begin{eqnarray}
   \label{InegraExpl}
   \int_0^\infty dx e^{-bx}x^l L^l_s(\lambda x) L^l_{s^\prime}(\eta
   x) =\frac{\Gamma(s+s^\prime+l+1)}{s! s^\prime!}
   \frac{(b-\lambda)^s(b-\eta)^{s^\prime}}{b^{s+s^\prime+l+1}}\times
   \\ \nonumber
   F(-s^\prime,-s;-s-s^\prime-l,\frac{b(b-\lambda-\eta) }{(b-\lambda)
   (b-\eta)}),
\end{eqnarray}
where $\Gamma$ is the gamma function, and $F$ is the
hypergeometric function ${^2F_1}(a,b;c,z)$.

Thus,  $\Psi_{El}^{II}$ in Eq.~(\ref{SolII}) can be written in
terms of the basis of $\Psi_{El}^{I}$ in Eq.~(\ref{GenSolI}), by
means of Eqs.~(\ref{expansionWFbarrier}) and~(\ref{expansionWF}),
as
\begin{eqnarray}
   \label{SolIIRE}\Psi_{El}^{II}(r,\phi,z)= \sum_{n^\prime,s}
   C_{n^\prime}^{II}  a_{ s s^\prime}^l f_{B}^{ns}(r,\phi,0)
   e^{ik_{n^\prime} z}.
\end{eqnarray}
The equality of $\Psi_{El}^{I}$  and $\Psi_{El}^{II}$  at $z=0$
implies
\begin{eqnarray}
   \label{WFEquality}
   C_n^I&=&\sum_{n^\prime} a_{ s s^\prime}^l C_{n^\prime}^{II}, \ \ \
   n\neq m \\
   \label{WFEquality2}
   1+C^{I}_m&=& \sum_{n^\prime}a_{s_m s^\prime }^l C_{n^\prime}^{II},
\end{eqnarray}
where each $n$ mode   is matched (as $l$ is fixed, summing over
$n^{\prime}$ implies summing over $s^{\prime}$.) Notice that this
matching also implies  that  of all the transverse functions. We
also impose the equality of the derivatives across the surface,
namely,
\begin{eqnarray}
   \label{matchpart}
   \partial\Psi_{El}^{I}/\partial z=
   \partial\Psi_{El}^{II}/\partial z
\end{eqnarray}
at $z=0$, which implies
\begin{eqnarray}
   \label{WFEqualityDer}
   C_n^I k_n&=&\sum_{n^\prime} a_{ s s^\prime}^l
   C_{n^\prime}^{II}k_{n^\prime}, \ \ \ n\neq m \\
   k_m^i+C^{I}_m k_m=(-1+C^{I}_m)k_m&=& \sum_{n^\prime} a_{s_m
   s^\prime
   }^lC_{n^\prime}^{II}k_{n^\prime}, \
\label{WFEqualityDer2}
\end{eqnarray}
where we also take account of  the reflected wave $m$, with
$s_m=m-l$, that propagates with momentum $k_m$,   opposite to that
of the incident wave. We thus find an infinite system of
non-homogenous linear equations for $C_n^{I}$, $C_m^{I}$ and
$C_{n^\prime}^{II}$, representing reflection and transmission
amplitudes, respectively. Their actual solution is analyzed in
Section~\ref{IV}.  This method is generalized in the appendix for
magnetic fields with arbitrary direction. While  we have assumed
an incoming wave with definite quantum numbers, the solution is
general for any incoming wave function, as one only needs to
expand it in the basis of Eq.~(\ref{eigenfunctionsCompl}), and
solve for each component. Indeed, a  wave function
$|\Psi_0\rangle$ representing initial conditions can be expanded
in terms of the solution basis:
\begin{eqnarray}
   \label{WFEqualityDeexpand}
   |\Psi_0\rangle= \sum_{ s l k_z} \langle s l
    k_z|\Psi_0\rangle |s l k_z\rangle ,
\end{eqnarray}
where an integral over the momentum $ k_z$ is implicit.

\section{Truncated solutions}\label{IV}

\begin{figure}[t] 
\vspace{-0.8cm} {\centering
\resizebox*{0.8\textwidth}{0.4\textheight}
{\includegraphics{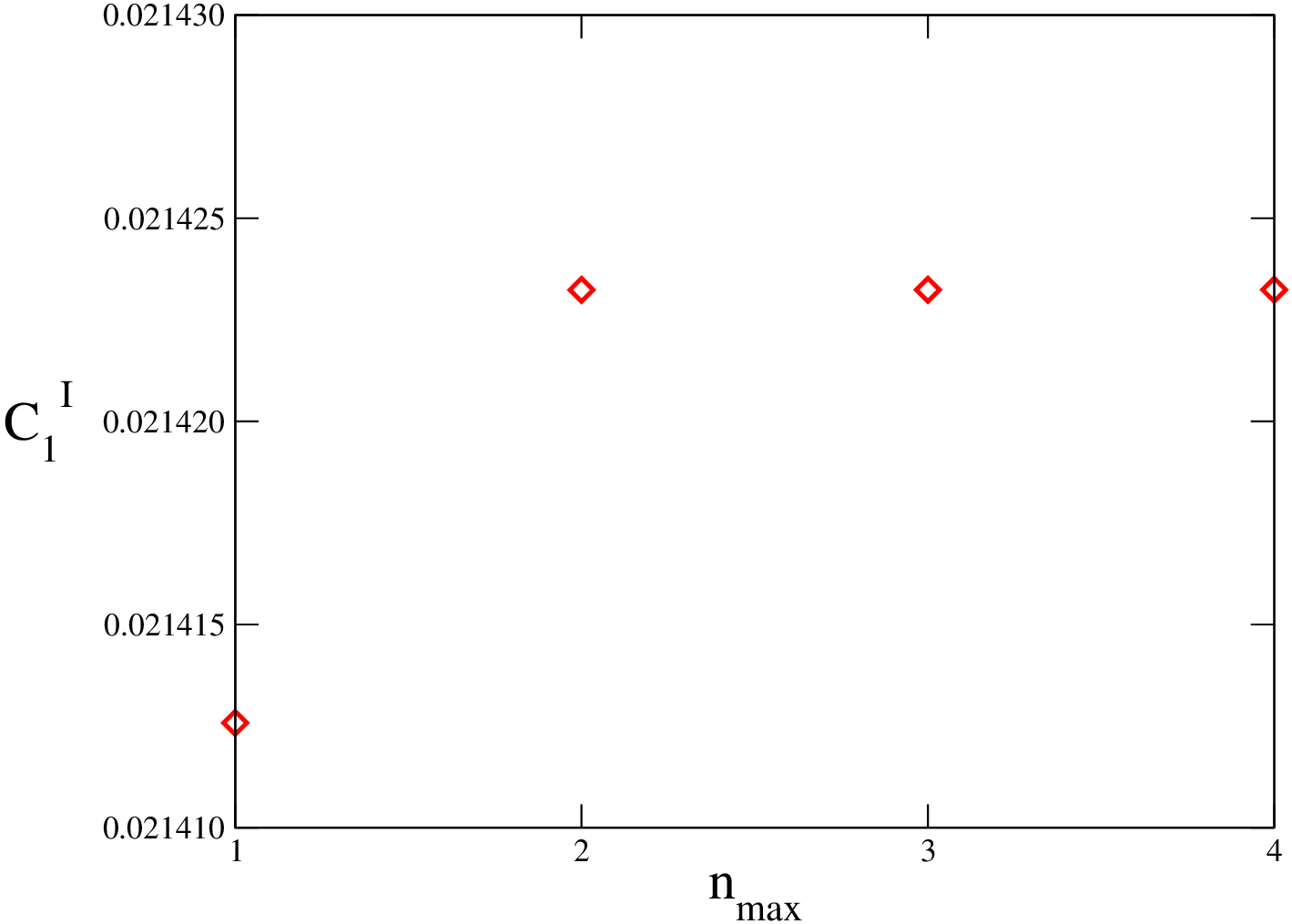}}\par} \caption{Sensitivity of
coefficient $C_1^I$ as a function of the maximum number of
coefficients in the truncated solution, $n_{max}$. The magnitude
of the hypermagnetic field in the symmetric phase is $B=0.01$
GeV$^2$. The magnitude of the magnetic field in the
broken-symmetry phase is $B'=\cos\theta_W B$. The incoming wave
function has quantum numbers $s=0$, $l=1$, an incident energy
$E=1.5$ GeV and mass $\mu=1$ GeV. The coupling constant in the
symmetric phase is taken as $e_0=g'=0.344$,  whereas in the
broken-symmetry phase it is
  $e_0'=e=g'\cos\theta_W$.} \label{fig1}
\end{figure}

In order to find the numerical solution to the system of
Eqs.~(\ref{WFEquality}),~(\ref{WFEquality2})
and~(\ref{WFEqualityDer}), (\ref{WFEqualityDer2}), we need to
truncate it. For an  incoming particle in region $I$ with
associated radial quantum number $m$, the relevant scattering
states into which it can be reflected are those with momentum
eigenvalue close to $k_m$, and therefore with radial quantum
numbers $n$ close to the incident one. For the transmitted states
on side $II$, it has been argued~\cite{AyaBespro} that for similar
values of $e_0B$ and $e_0^\prime B^\prime$ the main contribution
comes from the $m^\prime=m$ state and its closest neighbors.
Concentrating on such a case, it is sensible to formulate the
above system of equations considering the states from $m-n_{max}$
to $m+n_{max}$, and check the convergence for given integer
$n_{max}$. In fact, as the relevant $m$ is zero or close to it, we
shall consider $n\ \epsilon\ [0,n _{max}]$. Such a truncation is
appropriate since, as exemplified in Fig.~\ref{fig1} for
$C^{I}_1$, the coefficients quickly converge for a finite value of
$n_{max}$.

In such a case, the matrix representation of the above system of
equations can be given in terms of four blocks, each one of size
$n_{max}\times n_{max}$,
 \begin{eqnarray}
   \label{BlockMatrix}
   \left ( \begin{array}{cc}  \delta_{n^{\prime\prime} n}   &
   -a_{n^{\prime\prime} n^\prime}^l \\
   k_n\delta_{n^{\prime\prime} n} & -k_{n^\prime}  a_{n^{\prime\prime}
   n^\prime}^l \end{array} \right )
   \left (   \begin{array}{c} C_{n}^{I}  \\
   C_{n^\prime}^{II}   \end{array} \right ) = \left (   \begin{array}{c}
   -\delta_{ n^{\prime\prime }m}\\
   k_m \delta_{  n^{\prime\prime } m} \end{array} \right ),
\end{eqnarray}
where the upper blocks contain
Eqs.~(\ref{WFEquality}),~(\ref{WFEquality2}),
 and the lower ones Eqs. ~(\ref{WFEqualityDer}), (\ref{WFEqualityDer2})
 (the notation used keeps the $C_{n}^{I}$, $C_{n^\prime}^{II}$
indexes), and the terms with  $\delta_{ n^{\prime\prime }m}$
contain the non-homogenous part, and  define  the incoming $m$
state.

With the method thus described, we have obtained a quick
convergence of the reflection and transmission coefficients as
function of $n_{max}$, computed using a set of parameters
corresponding to the analysis in the following section. These
coefficients do not change after a relatively small value for
$n_{max}$ which for the chosen set of parameters corresponds to
$n_{max}\sim 20$. The sum of the reflection and transmission
coefficients adds up to unity. This point is further discussed in
the following sections.

\section{Transmission and reflection coefficients}\label{V}

To compute the reflection and transmission coefficients, we
consider the a flux of incoming positive-energy particles from
region $I$ with definite values of $k_m^i$ and quantum numbers $s$
and $l$. The conserved current is given by
\begin{eqnarray}
   \label{KleinGcurrent}
   j^\mu=\frac{-i}{2}\Psi^\dagger{\stackrel{\leftrightarrow}{\partial^\mu}}\Psi  -e_0\Psi^\dagger\Psi A^{\mu} .
\end{eqnarray}
We are interested in the longitudinal component of this current
integrated over the transverse $x-y$ plane at $z\rightarrow
-\infty$ for the incident and reflected waves and at $z\rightarrow
+\infty$ for the transmitted wave, namely
\begin{eqnarray}
   j=\int j^z dxdy,
   \label{integratedcurrent}
\end{eqnarray}
where $j$ represents the incident, reflected or transmitted
integrated currents
\begin{eqnarray}
   j _ i&=&k_m^i\\ \nonumber
   j _ r&=&\sum_n  {\rm Re}(k_n)     |C_{n}^{I}|^2 \\ \nonumber
   j _ t&=&\sum_{n^{\prime}}  {\rm Re}(k_{n^{\prime}})     |C_{n^{\prime}}^{II}|^2. \\
   \nonumber
   \label{currents}
\end{eqnarray}
The reflection and transmission coefficients $R$, $T$, are built
from the ratio of the reflected and transmitted integrated
currents along     the $\hat{z}$ direction, to the incident one:
\begin{eqnarray}
   \label{RefleTransmi}
   R&=&-j _ r/j _ i , \\
   T&=&j _ t/j _ i .
\end{eqnarray}
The incoming wave function corresponds to the first term on the
the right-hand side of Eq.~(\ref{GenSolI}), whereas the reflected
one corresponds to the second and third terms. The transmitted
wave function is given by Eq.~(\ref{SolIIRE}). The coefficients
$C_n^{I}$, $C_m^I$ and $C_{n^\prime}^{II}$ are found numerically
from the solution to Eqs.~(\ref{WFEquality}),~(\ref{WFEquality2})
and~(\ref{WFEqualityDer}), (\ref{WFEqualityDer2}) by truncating
the solution space as described in Sec.~\ref{IV}.

\begin{figure}[t] 
\vspace{-0.8cm} {\centering
\resizebox*{0.8\textwidth}{0.4\textheight}
{\includegraphics{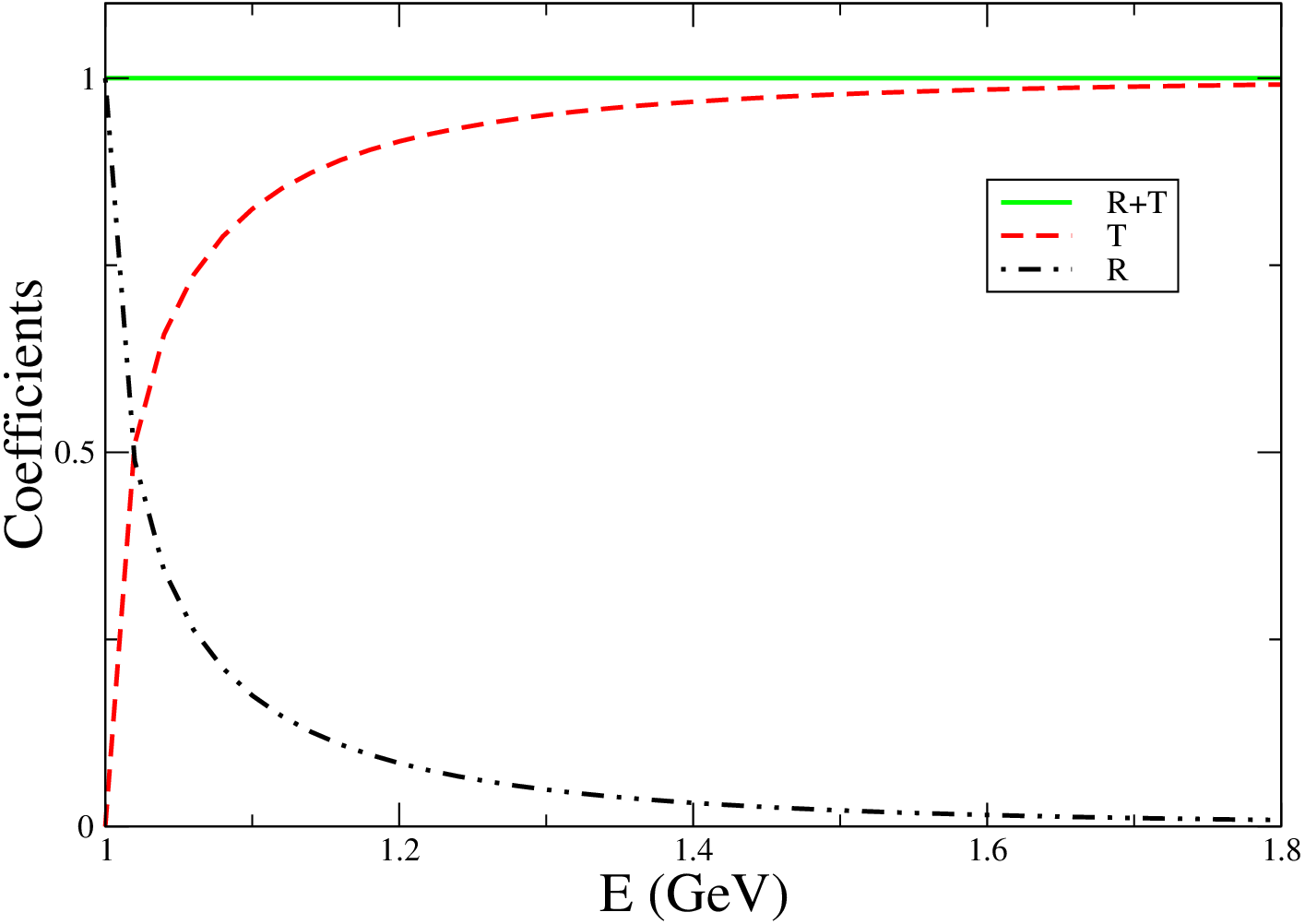}}\par} \caption{Reflection and
transmission coefficients for an incoming wave function with
quantum numbers $s=0$, $l=1$. The magnitude of the hypermagnetic
field in the symmetric phase is $B=0.01$ GeV$^2$. The magnitude of
the magnetic field in the broken-symmetry phase is
$B'=\cos\theta_W B$. The coupling constant in the symmetric phase
is   $e_0=g'=0.344$, whereas in the broken-symmetry phase it is
  $e_0'=e=g'\cos\theta_W$. The mass in the broken-symmetry
phase is   $\mu=1$ GeV. } \label{fig2}
\end{figure}

Figure~\ref{fig2} shows $R$ and $T$ for  the case of an incoming
wave function with quantum numbers $s=0$, $l=1$. The magnitude of
the hypermagnetic field in the symmetric phase is $B=0.01$
GeV$^2$. The magnitude of the magnetic field in the
broken-symmetry phase is obtained from the requirement that the
Gibbs free energy be a minimum, and is given by~\cite{Elmfors}
\begin{eqnarray}
   B'=\cos\theta_W B\, ,
   \label{B'andB}
\end{eqnarray}
where $\theta_W$ is Weinberg's angle. The coupling constant in the
symmetric phase is associated to the  hypercharge  $U(1)_Y$
symmetry, so that  $e_0=g'=0.344$. The coupling constant in the
broken-symmetry phase is associated to the electromagnetic
$U(1)_{em}$, that is, $e_0'=e=g'\cos\theta_W$. The mass in the
broken-symmetry phase is  $\mu=1$ GeV.

\begin{figure}[t] 
\vspace{-0.8cm} {\centering
\resizebox*{0.8\textwidth}{0.4\textheight}
{\includegraphics{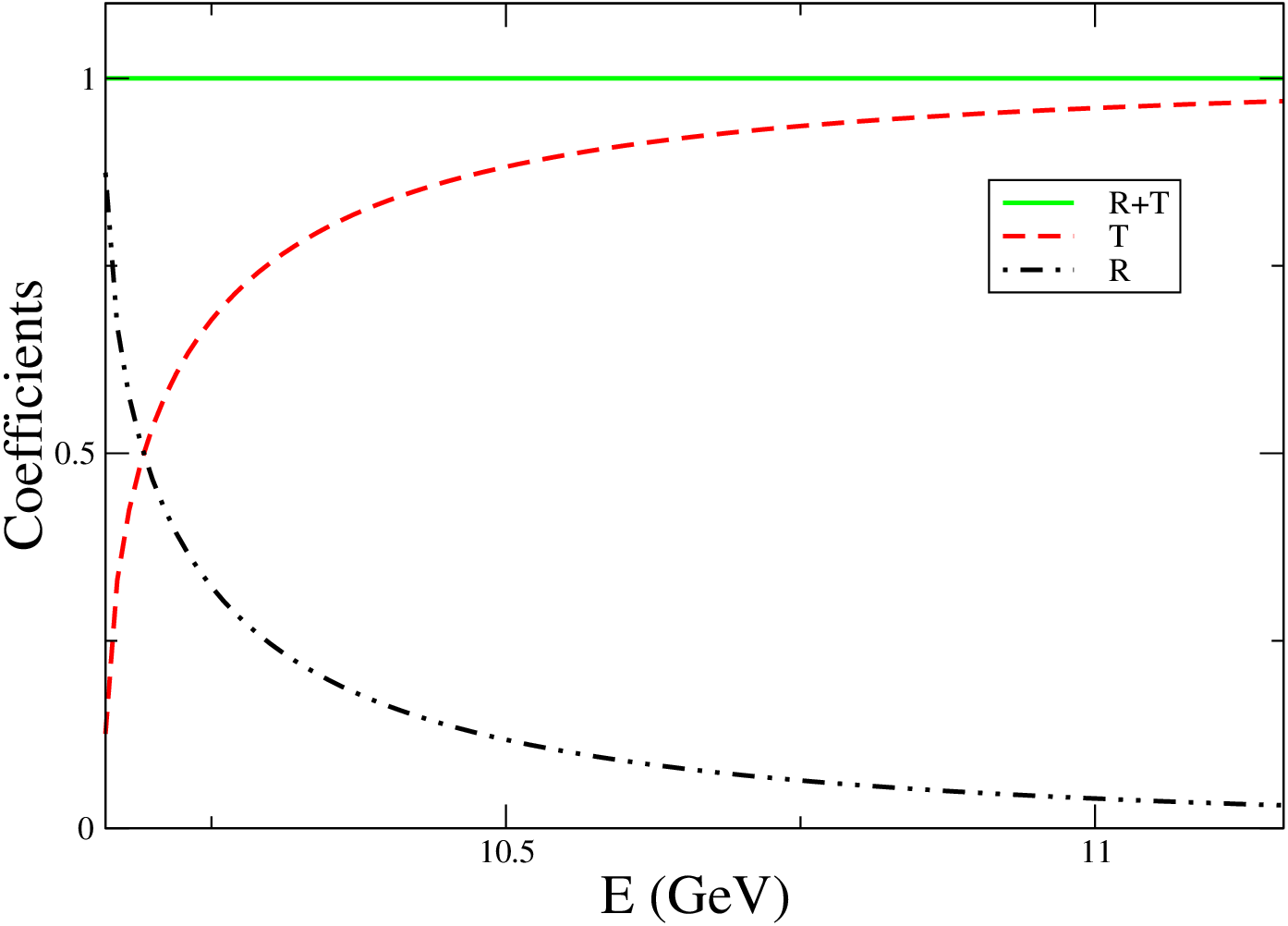}}\par} \caption{Reflection and
transmission coefficients for an incoming wave function with
quantum numbers $s=0$, $l=1$. The magnitude of the hypermagnetic
field in the symmetric phase is $B=100$ GeV$^2$. The magnitude of
the magnetic field in the broken-symmetry phase is
$B'=\cos\theta_W B$. The coupling constant in the symmetric phase
is $e_0=g'=0.344$, whereas in the broken-symmetry phase it is
$e_0'=e=g'\cos\theta_W$. The mass in the broken-symmetry phase is
$\mu=1$ GeV.} \label{fig3}
\end{figure}

Figure~\ref{fig3} also shows $R$ and $T$ for the same set of
parameters as for Fig.~\ref{fig2}, except that  the magnetic field
in the symmetric phase is  $B=100$ GeV$^2$.

Notice that in the cases depicted in Figs.~\ref{fig2}
and~\ref{fig3}, the  reflection and transmission coefficients add
up to unity, which in the present description of the scattering
problem corresponds to the conservation of incoming flux of
particles.

To understand such flux conservation, let us recall that,
according to Eq.~(\ref{massdecoupledeqs}), an incoming
positive-energy particle  from the symmetric phase encounters an
effective  potential barrier at the interphase of magnitude
\begin{eqnarray}
   \Delta V \sim \sqrt{2e_0'B^{\,\prime} (l+s'+1/2)+\mu^2} -
   \sqrt{2e_0B(l+s+1/2)}\, ,
   \label{potdiff}
\end{eqnarray}
which, for given quantum numbers of the incident flux, $l$ and
$s$, increases with  the radial quantum number $s'$ of the {\it
transmitted} wave. Indeed, a one-dimensional version of the
problem shows that such flux conservation, notwithstanding the
relativistic context, is related to a potential difference that is
mass-like. As argued below, this simulates the effect of a
non-relativistic Schr\"odinger equation. We also compare this
situation to the case of a gauge-like potential in the time
component.

\section{Toy models}\label{VI}

The conservation of flux can be understood with a simple
one-dimensional model where the wave function satisfies the
Klein-Gordon equation in two regions with different potentials.
With the same conventions for regions $I$ and $II$ as in
Sec.~\ref{III}, we consider
\begin{eqnarray}
   \label{KleinGFluxConse}
      [   \partial_t ^2-{\partial}_z^2 +V_1^2 ]
   \Psi & = & 0  ,  \ \ \  z < 0 \nonumber\\
 \left    [   \partial_t ^2-\partial _z^2  +V_2^2  + \mu^2 \right ]
      \Psi & =   & 0  , \ \ \  z > 0,
\end{eqnarray}
where $V_1$, $V_2$ are constant potentials for regions $I$ and
$II$, respectively. For $E>V_1$ and  $E>\sqrt{V_2^2+\mu^2}$ we
have propagating incident, reflected, and transmitted momenta
$k_i=\sqrt{E^2-V_1^2}$, $k_r=-k_i$, $q_t=\sqrt{ E^2-V_2^2
-\mu^2}$, respectively.
\begin{figure}[t] 
\vspace{-0.8cm} {\centering
\resizebox*{0.8\textwidth}{0.4\textheight}
{\includegraphics{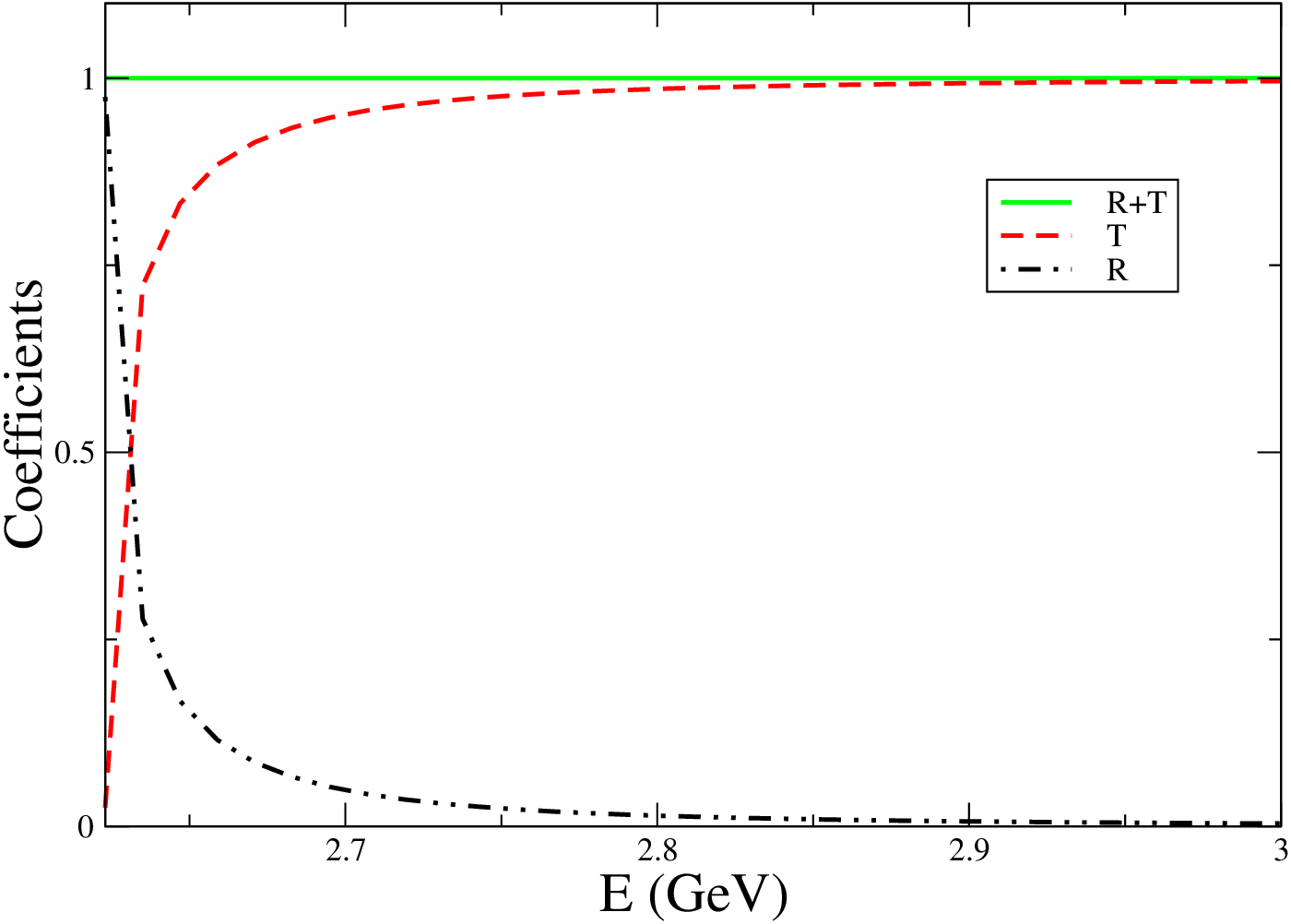}}\par} \caption{Reflection and
transmission coefficients in the one-dimensional with {\it
mass-like} one-dimensional potential in Eq.~(\ref{KleinGFlux}),
with $V_1=2.6$ GeV in region $z<0$, and $V_2=2.3$ GeV in region
$z>0$. The reflection and transmission coefficients add to one.
The mass in the $z>0$ region is $\mu=1$ GeV. }\label{fig4}
\end{figure}
Using the explicit form for the conserved current given in
Eq.~(\ref{KleinGcurrent}), and assuming an incoming flux of
positive-energy particles from region $I$, the reflection and
transmission coefficients are given by
\begin{eqnarray}
   R&=&\frac{(k_i-q_t)(k_i-q_t)^*}{ (k_i+q_t)(k_i+q_t)^*},\nonumber\\
   T&=&\frac{4 k_i^2}{ (k_i+q_t)(k_i+q_t)^*}\ ;
   \label{RTtoy}
\end{eqnarray}
these  correspond to the ratio of the reflected ($r$) and
transmitted ($t$) currents to the incident one, along the
$\hat{z}$ axis
\begin{eqnarray}
   \label{GcurrentratiosCon}
   j_{zr}/j_{zi}&=& -R\nonumber\\
   j_{zt}/j_{zi}&=& T \ {\rm Re} (q_t)/k_i ,
\end{eqnarray}
where the normalization is taken to the same density  $j_0$ as in
Eqs.~(\ref{KleinGcurrent}) for all cases. As shown in
Fig.~\ref{fig4}, we get current conservation in this case, which
we associate with the Schr\"odinger form of
Eq.~(\ref{KleinGFluxConse}).
\begin{figure}[t] 
\vspace{-0.8cm} {\centering
\resizebox*{0.8\textwidth}{0.4\textheight}
{\includegraphics{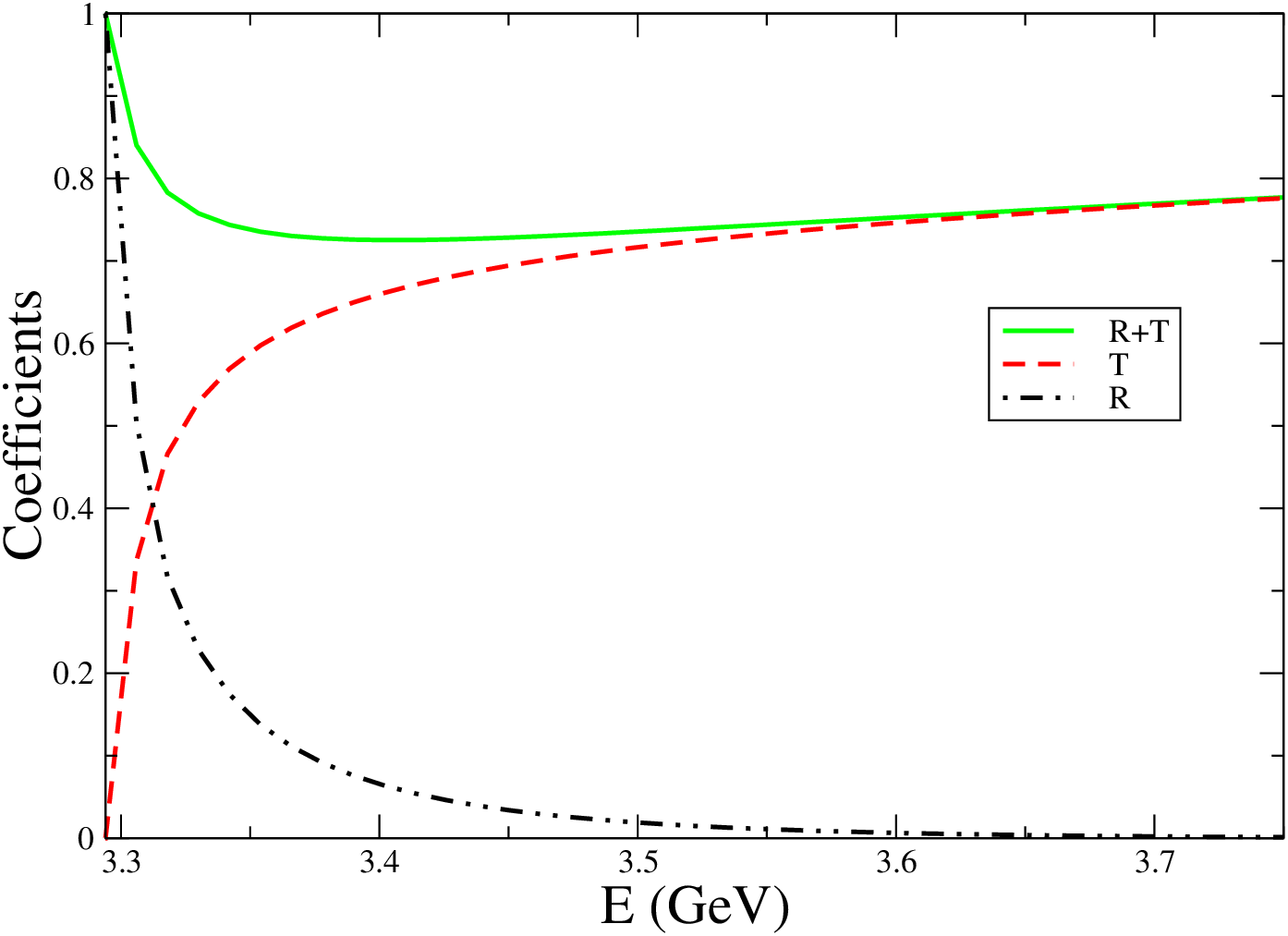}}\par} \caption{Reflection and
transmission coefficients with {\it gauge-like} one-dimensional
potential in Eq.~(\ref{KleinGFluxConse}), and $V_1=2.6$ GeV in
region $z<0$, and $V_2=2.3$ GeV in region $z>0$. The reflection
and transmission coefficients add to one only in the
non-relativistic and ultra-relativistic limits. The mass in the
$z>0$ region is $\mu=1$ GeV. }\label{fig5}
\end{figure}
This is not the case  for a gauge-like potential in the time
derivative, as in
\begin{eqnarray}
      [  (i\partial_t-V_1)^2-{\partial}_z^2 ]
   \Psi & = &  0  ,  \ \ \  z < 0 \nonumber\\
   \left    [  (i\partial_t-V_2)^2-\partial _z^2   + \mu^2 \right ]
      \Psi & =   & 0  , \ \ \  z > 0,
   \label{KleinGFlux}
\end{eqnarray}
where $V_1$, $V_2$ are constant potentials for regions $I$ and
$II$, respectively. For $E>V_1$ and  $E>V_2+\mu$ we have
propagating incident, reflected, and transmitted momenta
$k_i=E-V_1$, $k_r=-k_i$, $q_t=\sqrt{(E-V_2)^2+\mu^2}$,
respectively. The currents are
\begin{eqnarray}
   \label{Gcurrentratios}
   j_{zr}/j_{zi}&=& -R\nonumber\\
   j_{zt}/j_{zi}&=& \frac{E-V_1}{E-V_2}T \ {\rm Re} (q_t)/k_i ,
\end{eqnarray}
where $T$, $R$ are still those of Eqs.~(\ref{RTtoy}). It follows
that within the positive-energy solution approximation used, the
flux conservation $j_{xi}=j_{xr}+j_{xt}$ is valid   only in the
non-relativistic limit, and the ultra-relativistic limit
$E\rightarrow \infty$, as shown in Fig.~\ref{fig5}.  The flux
non-conservation in the intermediate regime  is associated to
particle-pair creation. The trouble comes from the relativistic
normalization in Eq.~(\ref{Gcurrentratios}). In practice, this
means that a particle encounters a potential that makes possible
the opening of additional channels contributing to the scattering
processes, such as pair production.

Such anomalous behavior is already known to arise in the context
of the Dirac equation in the treatment of scattering and is called
{\it Klein paradox}~\cite{Klein}. This signals the need to account
for negative-energy solutions. In fact, the gauge-like potential
in Eqs.~(\ref{KleinGFlux}) and the current in
Eq.~(\ref{KleinGcurrent}) produce the different current terms in
Eq.~(\ref{Gcurrentratios}). It follows that the
hypermagnetic-field problem resembles the first,
Schr\"odinger-like, case.

\section{Conclusions}\label{VII}

In this work, we  presented the exact solution for the scattering
problem of scalar particles propagating within two regions of
different  but constant  magnetic fields, separated by a planar
interphase. The fields are assumed to be perpendicular to the
interphase, while the solution for arbitrary directions is briefly
explained in the Appendix. The two regions are defined by the
condition that the particle is massless on one side of the
interphase and  massive on the other, such as during a first-order
EWPT, outside and inside the nucleated bubbles, respectively.

In the general case in which the direction of the magnetic field
is not perpendicular to the wall interphase, we can decompose it
into normal and parallel components. Recall that the normal
component of the external field is continuous across the
interphase and also that charged-particle trajectories consist of
an overall displacement along the field lines superimposed to
circular motion around the field lines. The relevant physical
effect of the magnetic-field component directed along the wall is
its ability to capture low momentum particles upon scattering
since these can make transitions to states that can remain
orbiting filed lines in this direction close to the wall.

In order to estimate the magnitude of such fields able to capture
particles on the wall, let us look at the following classical
argument: Equilibrium between the Lorentz force and the
centrifugal force gives for the radius of the orbit for a particle
trapped in the wall $R=p/(eB)$ where $p$ is the particle's
momentum, $e$ is its charge and $B$ the strength of the magnetic
field. Considering the physical situation of a wall with a finite
thickness $\lambda$, let us take $R\sim\lambda$. Also, let us
parameterize the strength of the magnetic field as $B=bT^2$, where
$b$ is a dimensionless factor and $T$ is the temperature, then
$b\sim p/(e\lambda T^2) $. Taking the momentum of a typical
particle to be of order of the temperature $T$, namely $p\sim T$
and since $\lambda\sim T^{-1}$ (see for example
Ref.~\cite{Ayala}), we obtain $b\sim {\mathcal{O}} (10) $ which is
already a very high value of the magnetic field. Although
uncertainties on the strength of  the magnetic fields at the EWPT
are large, it is safe to assume that $b\ll 10$ (see for example
Ref.~\cite{Elmfors}, and references therein).

We see that the loss of flux in these trapped modes for small
field strengths is small, and therefore that the relevant
magnetic-field component for the computation of transmission and
reflection coefficients is the normal one, thus the magnetic field
chosen in this paper. It is also worth mentioning  that for a
wall with non-zero thickness, realistic boundary conditions for
the external gauge field imply the existence of a magnetic-field
component along the direction of the wall and localized within the
dimensions of the wall (see the discussion in Sec. IV of the
second of Ref.~\cite{Ayala2}). In this case, similar arguments
apply and the conclusion is that only very intense fields are able
to produce an effect that involves a component of the magnetic
field parallel to the wall.

The problem at hand is solved through continuity conditions on a
surface, rather than a point.  The solution involves the
expression of the wave function on one side of the interphase as a
superposition of the complete set of solutions on the other side,
and the continuity of the wave function and its derivative across
the interphase. In order to numerically solve the problem, we
resorted to a truncation of the system of equations involving the
coefficients of the expansion.

We found that in such a scheme, the flux of particles is
conserved, which we associate with the Schr\"odinger nature of the
magnetic-field equation. We showed in a simple one-dimensional
model that such flux conservation can be understood as arising
from the contribution of an effective mass-like potential barrier
associated to momentum and radial modes, while,  in the case of a
gauge-like potential in the time component, the Klein
paradox~\cite{Klein}, signals the need to account for
negative-energy solutions.

As previously found~\cite{Ayala2, AyaBespro}, the  scattering of
chiral fermions during a first-order EWPT gives rise to an axial
asymmetry that can be subsequently converted into a baryon
asymmetry. The description of the scattering process used
simplifying assumptions for the reflected and transmitted states.
It will thus be interesting to explore the consequences of the
exact method introduced in this work in the context of the Dirac
equation on the generated axial asymmetry. This is work under
progress, and will be reported as a sequel of the present one.

\section*{Acknowledgments}

Support for this work has been received in part by DGAPA-UNAM
under grant numbers  IN108001, IN107105-3, and IN120602, and by
CONACyT-M\'exico under grant numbers 40025-F, and 42026-F.

\vskip .5 cm

\appendix{\Large \bf\noindent  Appendix: Solution for arbitrary
  magnetic fields}\label{IIIa} \vskip .5 cm

The method described in the paper can applied for the general case
of two regions separated by a wall with  fields ${\bf B}_1$,
${\bf B}_{2} $ that are neither necessarily parallel to each
other, nor perpendicular to the wall. Now we sketch the solution
to the same scattering problem as above.

We put the coordinates $(x,y,z)$ so that the wall separating the
two regions is on the $xz$ plane, so that $y<0$ for region $I$,
and $y>0$ for region $II$ (unlike the  above case.) The fields
define the plane $  B_1  B_2 $ perpendicular to the ${\bf
B_1}\times {\bf B_{2}}$ direction. We  also choose the $ \hat {\bf
z}$ axis along the line of the  $xz$ (wall)  and $B_1 B_2 $ planes
intersection. The inclination of plane $  B_1  B_2 $  with respect
to plane $xz$ is measured by an  angle $\phi$ drawn  by rotation
over  the  $ \hat {\bf z}$ axis. The directions  of the ${\bf
B}_1$, ${\bf B}_2 $ fields  on the $B_1  B_{2} $ plane (which by
definition lie on it) are measured  by angles    $\theta$ and
$\theta^\prime$ between the $ \hat {\bf z}$ axis and the $\hat{\bf
B}_1$, and $\hat {\bf B}_{2} $ directions, respectively. Indeed,
the system is  such that the spherical coordinates
$(B_1,\theta,\phi)$ and $(B_{2},\theta^\prime,\phi)$  also define
the fields. Actually, each field defines another   system of
cylindrical  coordinates such that travelling waves can be simply
described. These are $(r_1,\phi_1,z_1 )$ and   $(r_2,\phi_2,z_2
)$, where the $\hat {\bf z}_1$, $\hat {\bf z}_2$ axes are parallel
to ${\bf B_1}$, ${\bf B_2} $, respectively.

Assuming an incoming wave $f_{B}^{ns}(r_1,\phi_1,z_1)$ as in Eq.
(\ref{expansionWFnonPrime}), it will  be  reflected to
\begin{eqnarray}
   \Psi_{E }^{I}(r_1,\phi_1,z_1)=  \sum_{nl} C_{nl}
   f_{B_1}^{ns}(r_1,\phi_1,z_1)
   \label{firsttolast}
\end{eqnarray}
(note that for didactic purposes, the arrangement is different
than in Eq. (\ref{GenSolI}),) and transmitted  to
\begin{eqnarray}
   \Psi_{E }^{II}(r_2,\phi_2,z_2) =\sum_{n^\prime
   l^\prime} C_{n^\prime l^\prime} f_{B_2}^{n^\prime
   s^\prime}(r_2,\phi_2,z_2),
   \label{last}
\end{eqnarray}
respectively, on each side of the wall. Here we  also sum  over
the angular momentum  quantum number, which  is no longer
conserved, while the energy is. The continuity of the wave
function on the $xz$ plane requires calculation of the overlap of
the  wave-function components. This is carried out transforming
the coordinates  $(r_1(x,y,z) ,\phi_1(x,y,z),z_1(x,y,z) )$  and
$(r_2(x,y,z),\phi_2(x,y,z),z_2(x,y,z) )$ that are obtained
respectively by the rotations. The condition for the wave function
on the wall is $y=0$, so that the overlapping amplitudes  as in
Eq.~\ref{amplitudBBp} can be calculated. We find that  evanescent
waves also play an important role as they provide the complete
basis that allows  for boundary conditions to be satisfied on a
surface.

As for the  case treated in the paper, one can use this method for
an arbitrary incoming  function, expanding it in the  above basis.


\end{document}